# The investigation of EPR paramagnetic probe line width and shape temperature dependence in high-temperature superconductors of Bi-Pb-Sr-Ca-Cu-O system.

J.G.Chigvinadze[a], J.V.Acrivos[b], I.G.Akhvlediani[a], M.I.Chubabria[c], T.L.Kalabegishvili[a], T.I.Sanadze[d].

a – Andronikashvili Institute of Physics, 0177 Tbilisi, Georgia, Tamarashvili st. 6, Tel.: 995 32 397924, Fax: 995 32 391494, jaba@iphac.ge.
b – San Jose State University, San Jose CA 95192-0101, USA, Tel.: 408 924 4972, Fax: 408 924 4045, jacrivos@athens.sjsu.edu.
c – Institute of Cybernetics, 0186 Tbilisi, Georgia.
d – Javakhishvili State University, 0178, Tbilisi, Georgia.

Abstract

The work is related with the finding out of magnetic phases in strongly anisotropic high-temperature superconductor $Bi_{1.7}Pb_{0.3}Sr_2Ca_2Cu_3O_{10-\delta}$ in the temperature region where the superconductor is in the normal state. It was studied the temperature dependence of the paramagnetic probe EPR line width. In the normal state at $T>T_c$ near 175 K it was revealed a pick in the temperature dependence of line width. In this region it was observed the time increase of the line width with the characteristic time ~ 17 min. This shows the possibility of magnetic phase formation in this material.

## 1. Introduction

The Bi-Pb-Sr-Ca-Cu-O system is one of the most perspective material from the point of view of technical application of high-temperature superconductivity (HTSC) [1]. It is characterized by a high critical temperature of superconductive transition $T_c$=107K and a high upper critical magnetic field $H_{c2}$ of the order of 150T [2]. The contemporary technology of superconductor's synthesis makes it possible the changing of their critical current density $J_c$ [3]. Its high value is necessary also for applications of high-temperature superconductors in one of the most perspective direction of contemporary technique – strong-current energetics [4].

The Bi-Pb-Sr-Ca-Cu-O system is characterized by a such high critical temperature of superconducting transition $T_c$ that it remains superconductive also at temperatures when thermal fluctuations play significant role because their energy becomes comparative with the elastic energy of vortices and also with the pinning energy [5]. This creates prerequisites for phase transitions. Due to the layered crystal structure and anisotropy of HTSC it appears conditions for creation of different phases in them [6-17], as example, in the Bi-Pb-Sr-Ca-Cu-O(2223) system the three-dimensional 3D Abrikosov vortices experience a phase transition in the quasi-two-dimensional 2D vortices, so-called "pancake"-ones, at increasing the magnetic field(in fields of the order of several kOe).

At investigations of dissipations processes in HTSC it was revealed that the energy absorption of low-frequency oscillations was sharply reduced and fell down approximately on two orders of value [14,15]. The reason for a such sharp reducing of energy absorption of low-frequency oscillation is a step-wise increase of pinning force anticipated by the American theoreticians [16] and experimentally observed in work [17]. It is known that the phase transition (3D-2D) in a high-temperature vortex matter is caused by its layered crystal structure and strong anisotropy (the anisotropy factor of this superconducting system is of the order of 3000) [14]. Another example of phase transition in the vortex matter of HTSC is the Abrikosov vortex lattice melting near $T_c$. During the Abrikosov vortex lattice melting it is sharply changed the character of relaxation phenomena.

At temperatures much lower $T_c$ it is observed the slow logarithmical reduction of captured magnetic flux [18-20] what is explained by the Anderson creep [21]. Near $T_c$ in the region of Abrikosov vortex melting temperature the logarithmic character of relaxation is changed by the power dependence with exponent 2/3 [22].

Consequently, the investigation of phase transitions in HTSC is important for the understanding of processes taking place in these materials.

It is important also the finding out of new HTSC phases with higher transitions temperatures $T_c$ in the superconducting state. It should be noted that for the understanding of processes observed in HTSC it could be decisive the study of processes taking place in superconducting samples in their normal state, i.e. at temperatures higher then $T_c$.

## 2. Experiment

For investigations of superconductive transitions [23] and phase transitions in the vortex matter, and also for the study of magnetic flux structure [24,25], one could use the highly sensitive and informative electron paramagnetic resonance method. For this aim it is necessary to coat the surface of sample by a paramagnetic probe, as example, by the diphenylpicrylhydrazil (DPPH) and to record the EPR line width and amplitude of this radical at the temperature change both above and below the critical temperature of superconducting transition.

A great importance could be also the microwave nonresonant absorption using an EPR spectrometer.

For the first time the EPR probe investigations of superconductors was presented in work by B.Rakvin et al. [26] in 1989. It is based on the earlier work by P.Pincus et al. [27] where using the NMR method it was observed the line broadening at superconducting transition in type II superconductors. The line broadening take place due to a magnetic field inhomogeneity at appearance of Abrikosov vortices [28] in the mixed state, when $H_o$ is in the range between the first and the second critical magnetic fields: $H_{c1}<H_o<H_{c2}$.

In the case of electron paramagnetic resonance it is changed the EPR line width of paramagnetic DPPH probe coating on the surface of a sample. The magnetic field inhomogeneity caused by the Abrikosov vortex lattice does not depend on the field in the range of $H_{c1}<H_o<H_{c2}$ and the second moment of a field distribution is expressed by [27]:

$$\overline{\Delta H_c^2} = \frac{\Phi_0^2}{16\pi^3\lambda^4} \quad (1)$$

where $\Phi_o$ is the flux quantum ($\Phi_o=2.07\cdot10^{-7}$ G·cm$^2$) and $\lambda$ is the field penetration depth. If the second moment of the EPR line observed just above $T_c$ is denoted by $\overline{\Delta H_n^2}$, the total measured peak-to-peak line width is given by

$$\Delta H_{pp} = 2(\overline{\Delta H_n^2} + \overline{\Delta H_s^2})^{1/2} \quad (2)$$

The $\overline{\Delta H_n^2}$ is determined by the spin interactions of DPPH radicals coating the surface and does not change when the sample becomes superconducting.

Investigating the temperature dependence of EPR paramagnetic probe line width one could observe different magnetic transitions [29]. For this aim our $Bi_{1.7}Pb_{0.3}Sr_2Ca_2Cu_3O_{10-\delta}$ sample was coated by a stable DPPH radical. The DPPH probe was coated on the surface of a sample. The latter was immersed in the $10^{-2}$M solution of DPPH in acetone and then dried.

For the temperature measurement of EPR line width, the double quartz ampule with a sample, heating unit and thermocouple were placed into quartz cryostat with liquid nitrogen.

The EPR measurements were carried out on the RE1306 spectrometer operating at 9,4 GHz (X-range) frequency and 100 kHz modulation. The amplitude and level of microwave power were chosen in the way to exclude the broadening and saturation of the EPR line.

The EPR installation was operated in the temperature range 77-300K.

The heating unit was wound on the inner ampule so that to exclude the reduction of resonator's quality factor [30]. Changing the current value through the heating unit one could reach necessary temperature in a sample. The temperature stability was ~0,2K.

### 3. Results and their discussions.

The presented work is devoted to the finding out the magnetic phases in HTSC in temperature range above $T_c$=107K up to 300K, i.e. in the temperature region where the high-temperature sample is in the normal state.

Temperature measurements were carried out in the following way: a sample, placed in a magnetic field was quickly cooled down to 77K transforming it into superconducting state. Then the line width and line amplitude was measured, firstly at increasing temperature from 77K up to 300K, and then decreasing it from 300K down to 77K.

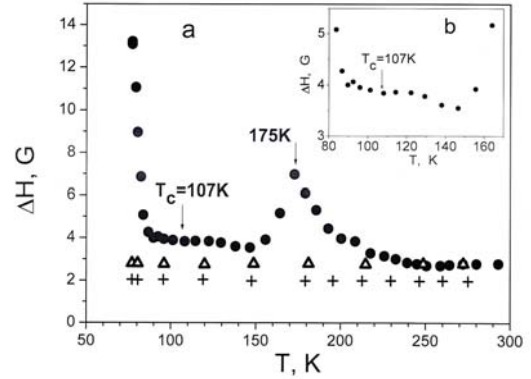

Fig. 1. The EPR line width dependence on temperature:
● - for DPPH adsorbed on the surface of $Bi_{1.7}Pb_{0.3}Sr_2Ca_2Cu_3O_{10-\delta}$.
△ - for DPPH adsorbed on the surface of copper
+ - for the powder-like DPPH.

In Fig.1 it is presented a temperature dependence of EPR line width of the paramagnetic probe coated on the surface of HTSC $Bi_{1.7}Pb_{0.3}Sr_2Ca_2Cu_3O_{10-\delta}$(2223). As it is seen from the figure, that in the region of 175K temperature it is observed the clear-cut pick of the EPR line width.

On the right side of the observed pick up to room temperature and on the left side from the pick in the temperature interval from 150K to 107K the EPR line width does not depend on temperature, meanwhile at the pick temperature (Fig.2) the line width changes in time and is described by the expression:

$$\Delta H_{pp} = \Delta H_{pp}^0 - A\exp(-t/t_1) \quad (3)$$

where $\Delta H_{pp}^0 = 4.2 \pm 0.08$G; $A = 1.18 \pm 0.07$G; $t_1 = 17 \pm 2.6$ min.,

i.e. at this temperature, from our point of view, it takes place the formation of magnetic phase with a characteristic time ~ 17 min.

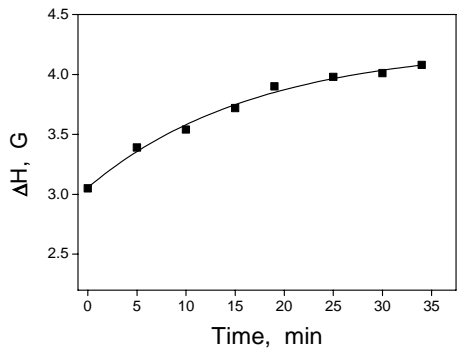

Fig. 2. The time dependence of the EPR paramagnetic probe line width of the $Bi_{1.7}Pb_{0.3}Sr_2Ca_2Cu_3O_{10-\delta}$ sample at 175 K temperature.
■ – experiment;
solid curve – the approximation by exponent.

The increase of line width, by the way essential, is observed below $T_c$ what is connected with the transition of a sample in the superconductive state during which the sample is penetrated by Abrikosov vortices [28] changing sharply the magnetic field homogeneity what is related with the EPR line width increase.

The test experiments carried out on a powder-like DPPH and also on a DPPH coating a normal metal copper sample the same size and shape as the $Bi_{1.7}Pb_{0.3}Sr_2Ca_2Cu_3O_{10-\delta}$ sample under investigation, showed the absence of the EPR line width change in the whole temperature interval 77-300K (Fig.1). So, it is seen that the pick at 175K is related only with the HTSC $Bi_{1.7}Pb_{0.3}Sr_2Ca_2Cu_3O_{10-\delta}$ sample being in the normal state.

To clear out the fact whether this pick is related with superconductivity we transformed our sample in non-superconductive state. For this aim we heated it up to the melting temperature and then quickly cooled it down to room temperature. After such treatment the sample was again coated by the DPPH layer. The experiment showed the absence of the EPR line broadening below 107K, i.e. below the critical temperature of superconducting transition at the decrease of temperature, the EPR line broadening was not observed but the pick in the high-temperature region remained, although it was decreased in the value (Fig.3)

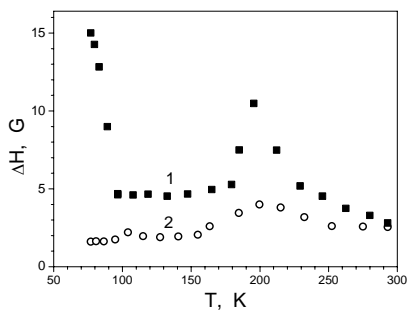

Fig.3. The temperature dependence of EPR paramagnetic probe line width of the $Bi_{1.7}Pb_{0.3}Sr_2Ca_2Cu_3O_{10-\delta}$ sample,
1 - before a thermal treatment
2 - melted and quickly cooled down to room temperature.

The EPR line width pick of high-temperature Bi-Pb-Sr-Ca-Cu-O(2223) superconductor observed by us is apparently connected with the essential change of material's properties (as example, with the formation of a new magnetic phase with completely unusual characteristics), because along with the EPR line width pick appearance with the maximum at 175K in this temperature region, it is observed an unusual temperature dependence of the EPR signal amplitude (Fig.4).

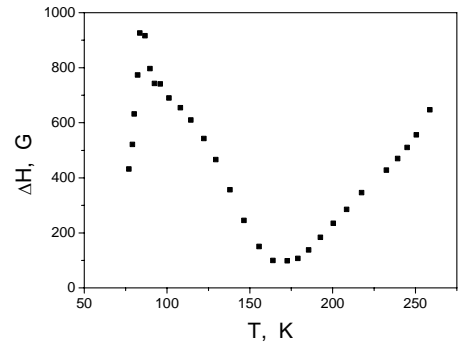

Fig.4. The temperature dependence of the EPR line amplitude of the $Bi_{1.7}Pb_{0.3}Sr_2Ca_2Cu_3O_{10-\delta}$ system.

It is seen from the figure that in the temperature interval where the sample is in the superconducting state with the decrease of EPR line width the signal amplitude increases what should be observed. But after the transition in the normal state the line width in the temperature interval 107-150 K remains constant (see Fig.1) but the amplitude decreases more sharply then it is expected from the Curie law. The observed change of signal amplitude could be caused by the change of signal shape. It should be noted that the signal amplitude minimum is reached at temperature where it is observed the EPR line width maximum (175K).

And, finally, it is should be noted the hysteresis phenomena observed in the temperature dependence of the EPR line width.

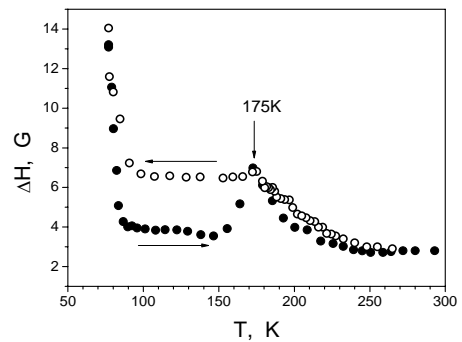

Fig.5. The hysteresis curve of the EPR line width temperature dependence of the paramagnetic probe in high-temperature $Bi_{1.7}Pb_{0.3}Sr_2Ca_2Cu_3O_{10-\delta}$ superconductor.

As it is seen from Fig.5, at the decrease of temperature (300→77K) at 175K the line width pick is not observed. So, in the formation of magnetic phase the important role is played by the sample prehistory, i.e. whether the 175K temperature is approached from the superconductive or the normal states.

## 4. Conclusions.

1. It was observed the pick in the EPR line width temperature dependence in high-temperature samples of $Bi_{1.7}Pb_{0.3}Sr_2Ca_2Cu_3O_{10-\delta}$ system in the normal state, in the region of 175K temperature.

2. It was shown that in the temperature region of pick (T~ 175 K) the EPR line width of paramagnetic probe increases with time and is described by the exponent with characteristic time ~ 17 min.

3. It was observed the anomaly in the temperature dependence of EPR line amplitude: the temperature of the line amplitude minimum coincides with the one of the line width maximum. The observed anomaly in the temperature dependence is apparently caused by the change of the EPR line shape.

4. It was observed the hysteresis phenomena in the temperature dependence of the EPR line width.

The results obtained make it possible to conclude that in the region of temperature near 175 K it is apparently formed a new magnetic phase.

### Acknowledgements

The work is supported by the grants of International Science and Technology Center (ISTC) G-389 and G-593.